\documentclass[cameraready]{Interspeech}

\title{Audio Imitator: Controlling Timbre and Tempo in Video2Audio Synthesis with Audio Reference}

\author[affiliation={1,2}, orcid=0000-0000-0000-0000]{Jiahui}{Zhao} 
\author[affiliation={1}, orcid=0000-0000-0000-1111]{Tianrui}{Wang}
\author[affiliation={1,2}]{Chunyu}{Qiang}
\author[affiliation={1}]{Cheng}{Gong}
\author[affiliation={2}]{Xijuan}{Zeng}
\author[affiliation={2}]{Feng}{Deng}
\author[affiliation={1}, correspondingauthor]{Longbiao}{Wang}

\address{
    $^1$ Tianjin University, China \\
    $^2$ Kuaishou Technology, China
}

\email{first@university.edu, second@companyA.com, third@companyB.ai}


\usepackage{comment}

\begin{document}

\maketitle

\begin{abstract}
    Video-to-audio generation has made significant progress in achieving semantic consistency and temporal alignment from silent videos. However, audio contains rich stylistic attributes such as timbre and tempo that are difficult to infer from visual and textual inputs alone. While reference audio can serve as additional conditioning, it is typically treated as a holistic signal, limiting fine-grained style control. We propose AudioIM, an attribute-aware framework that explicitly models timbre and tempo as separate control factors rather than relying on holistic prompt conditioning. Dual encoders extract complementary timbre-related and tempo-related representations, which are injected through global conditioning. A masking-based training strategy enables effective latent prompt conditioning at inference. Experiments on VGGSound show improved style similarity while preserving semantic alignment and synchronization. Audio samples are available at: \url{https://anonymousdemo757.github.io/}.
\end{abstract}

\section{Introduction}

Recent advances in Artificial Intelligence Generated Content (AIGC) \cite{you2025ta} have significantly accelerated progress in cross-modal generation tasks, including video-to-audio (V2A) synthesis. Generating temporally aligned and semantically consistent audio from silent videos has broad applications in virtual reality, automatic Foley production, and embodied perception systems \cite{wang2024tiva, xu2024video, ren2025sta, zhang2024foleycrafter, cheng2025mmaudio}. Existing works have demonstrated strong performance in aligning generated audio with visual events under textual guidance \cite{xu2024video, ren2025sta, zhang2024foleycrafter, cheng2025mmaudio}, but they generally condition only on visual/text content and barely explore the controllability of various audio properties.

Audio signals contain a wealth of information that cannot be fully described by video and text alone, such as timbre and tempo. Timbre refers to voiceprint-like characteristics (e.g., the individual identity that distinguishes one cat’s meow from another), while tempo refers to characteristics similar to music that determine the overall rhythm. To achieve control over the style of generated audio, using audio reference as prompt becomes inevitable.
Recent studies \cite{du2023conditional, chen2025video} have explored incorporating reference audio as an additional conditioning signal for V2A generation. In these approaches, the reference is typically introduced through latent concatenation or continuation-based strategies, enabling the model to transfer overall acoustic characteristics from the prompt audio. While effective, the reference audio is generally treated as a holistic conditional signal, where style-related attributes remain implicitly embedded within the latent representation. As a result, fine-grained control over specific style components remains limited.

In this work, we propose an attribute-aware and style-controllable V2A framework that models reference audio beyond holistic conditioning, improving generation diversity and enabling fine-grained control over audio style.
Inspired by zero-shot techniques in TTS \cite{le2023voicebox, li2024styletts, ju2024naturalspeech, yushenf5, zhou2025indextts2}, to enable effective utilization of prompt audio latents, we introduce a masking-based training strategy that allows the model to learn conditional flow matching with partially observed audio representations. This design enables the system to leverage reference audio for fine-grained style guidance during inference, while preserving semantic consistency and temporal alignment with the video. The prompt audio latent provided fine-grained temporal tempo and timbre information for model inference. 
Instead of treating prompt audio as a single undifferentiated signal, we introduce attribute-aware style modeling that separately captures timbre-related and tempo-related information. Specifically, we adopt dual encoders to extract complementary style representations from the reference audio: a timbre encoder to capture voiceprint-like characteristics, and a tempo encoder to capture global rhythmic structure. These style features are fused and injected into the generation backbone through global conditioning.

Our proposed AudioIM framework is built upon the MMAudio backbone and extends it with (i) masked latent prompting for style-aware conditioning and (ii) global style injection via dual encoders. Experimental results on the VGGSound benchmark demonstrate that our approach improves style similarity to reference audio while maintaining strong semantic alignment and synchronization performance.
In summary, our contributions are as follows. We introduce an attribute-aware style modeling framework for video-to-audio generation that structures reference audio conditioning into timbre and tempo components. We propose a masking-based training strategy that enables effective utilization of prompt audio latents during inference. We demonstrate improved controllability of audio style while preserving semantic consistency and temporal alignment.

\begin{figure*}[htbp]
    \noindent
    \includegraphics[width=0.95\textwidth]{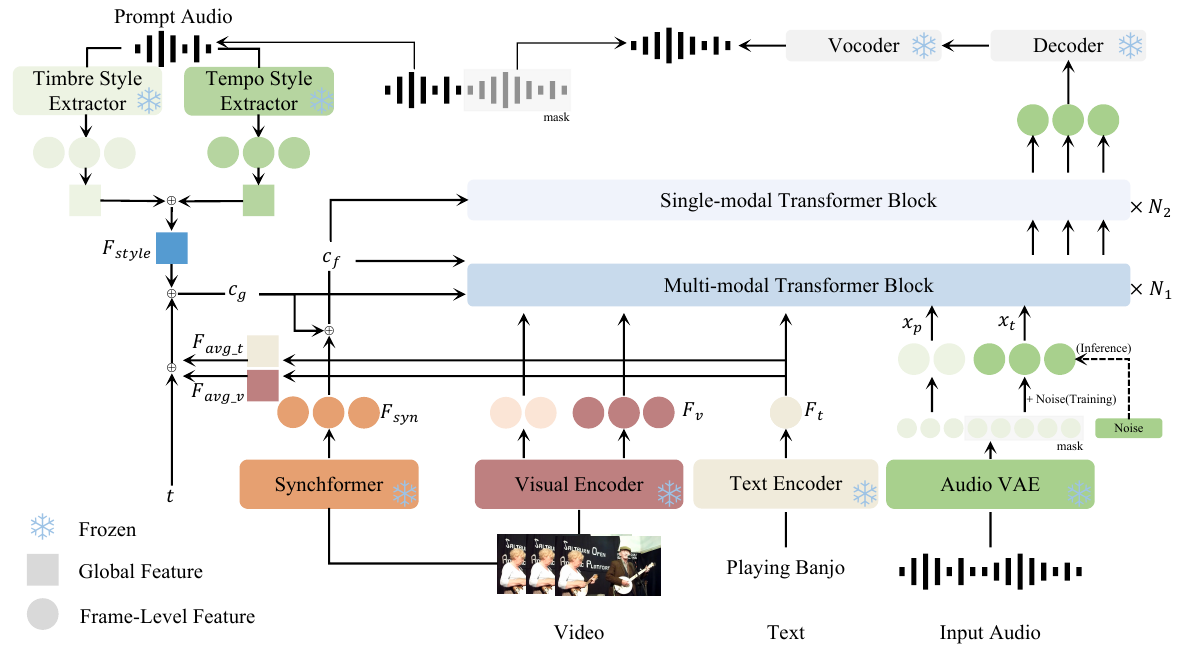}
    \vspace{-10pt}
    \caption{Overview of the proposed AudioIM framework. During training, audio latents are partially masked to enable latent-level prompt conditioning under a flow-matching objective. Reference audio is processed by dual style encoders to extract complementary timbre-related and tempo-related representations. These style features are fused and injected through global conditioning pathways, enabling style-aware video-to-audio generation while preserving semantic and temporal alignment.}
    \vspace{-10pt}
    \label{flowchart}
\end{figure*}

\section{Method}

Fig.~\ref{flowchart} illustrates the overall structure of our method.
We aim to leverage a reference audio to provide style information that cannot be directly inferred from video–text–audio triplets, where video and text lack explicit cues about audio style. In the audio style transfer task, it has been shown that sound style is composed of timbre and tempo. 
To this end, we introduce a masking strategy during training, where the latent representation of the reference audio supplies style information to the model, enabling zero-shot inference. 
In addition, we employ dual encoders to extract global style features from the reference audio, which are incorporated into the model as conditional guidance.

\subsection{Masking Training}

We apply masking training \cite{yushenf5, rouard2024audio} to the audio latents output by the audio variational auto-encoder (VAE) \cite{bengio2014auto}. Specifically, the audio input $x$ is encoded by the VAE and then masked with a 3:5 ratio. One part is used as the flow-matching conditional input $x_p$ to provide stylistic information, while the other part is input as $x_1$ and combined with noise $x_0$ during training. The input video $V$ is encoded by a video CLIP \cite{radford2021learning} encoder to obtain the video semantic representation $F_v$, and further processed by Synchformer \cite{iashin2024synchformer} to extract audio-visual alignment features $F_syn$, while the text $T$ is encoded by a text CLIP \cite{radford2021learning} encoder to obtain the text representation $F_t$.
\begin{equation}
    \begin{split}
    &\boldsymbol{x}_{p} = m \odot \text{VAE}(x), \boldsymbol{x}_{1} = (1-m) \odot \text{VAE}(x), \\
    &\boldsymbol{x}_{t} = t * \boldsymbol{x}_{1} + (1-t) * \boldsymbol{x}_{0}, \\
    &\boldsymbol{F}_{v} = \text{CLIP}(V), \boldsymbol{F}_{syn} = \text{CLIP}(V), \\
    &\boldsymbol{F}_{t} = \text{CLIP}(T),
    \end{split}
\end{equation}
Audio latent $x_p$, together with the video semantic latent $F_v$ and the text latent $F_t$, is integrated to construct the condition $C$, which serves as the condition of flow matching \cite{lipman2022flow}.
The model’s global conditioning $C_g$is constructed from three components: the timestep embedding $t$, the average-pooled video semantic representation $F_{avg}\_v$, and the average-pooled textual representation $F_{avg}\_t$. 
\begin{equation}
    \boldsymbol{C}_{g} = \text{MLP}(\boldsymbol{F}_{avg\_v} + \boldsymbol{F}_{avg\_t}) + \boldsymbol{t}
\end{equation}
The frame-level conditioning $C_f$ is obtained by upsampling the aligned feature $F_{syn}$.
\begin{equation}
    \boldsymbol{C}_{f} = \text{Upsample}(\text{ConvMLP}(F_{syn})) + \boldsymbol{C}_{g},
\end{equation}
The global and frame-level conditions are applied to the video-text and audio modalities via adaLN \cite{perez2018film}, respectively.
\begin{equation}
    \begin{split}
    &(\boldsymbol{x}_{p}^{'}, \boldsymbol{x}_{t}^{'}) = \text{adaLN}(\text{projector}(\boldsymbol{x}_{p}, \boldsymbol{x}_{t}), \boldsymbol{C}_{f}), \\
    &\boldsymbol{F}_{v}^{'} = \text{adaLN}(\text{projector}(\boldsymbol{F}_{v}), \boldsymbol{C}_{g}), \\
    &\boldsymbol{F}_{t}^{'} = \text{adaLN}(\text{projector}(\boldsymbol{F}_{t}), \boldsymbol{C}_{g}),
    \end{split}
\end{equation}
During training, the conditional flow matching objective is as follows:
\begin{equation}
    \mathcal{L}_{FM}(\theta) = \mathbf{E}_{t, q(x_0), q(x_1, C)} \lVert {v}_{\theta}(t, C, x_t) - u(x_t) \rVert^2.
\end{equation}

where $\theta$ denotes the network parameters, $v_{\theta}(t, C, x_t)$ represents a time-dependent conditional velocity field, $q(x_0)$ is the standard normal distribution, $q(x_1, \mathbf{C})$ samples from the training data, $u(x_t) = x_1 - x_0$ denotes the corresponding flow velocity at $x_t$.

Through mask training, the model is enabled to perform flow-matching inference conditioned on prompt audio latents, thereby leveraging them to provide fine-grained stylistic information for audio generation.

\subsection{Attribute-aware Style Condition}

To enhance style control in video to audio generation, We separately extract features containing timbre and tempo information from the prompt audio and model them independently. Specifically, we leverage dual encoders to extract features from the reference audio $x_{ref}$. The timbre encoder is implemented based on BEATs \cite{chen2022beats}. BEATs is a foundation audio representation model pre-trained on AudioSet, which converts continuous audio signals into semantically rich and abstract representations. BEATS achieves state-of-the-art performance on multiple audio and speech classification benchmarks, demonstrating strong generalization ability and robustness to acoustic variations, and is capable of extracting representations that contain voiceprint-like characteristics. The tempo encoder adopts the Style Conditioner proposed in \cite{rouard2024audio}, in which it is demonstrated that this module takes a few seconds of audio as input and extract features from it. The extracted features are used for high-level conditioning, such as tempo and harmony, while still retaining some lower-level content such as melodic patterns or rhythm. The conditioning module introduces a Residual Vector Quantizer based information bottleneck, ensuring that sufficient features are extracted for meaningful conditioning, but not so much that the generative model simply copies or loops the reference audio. We employs six codebooks to extract richer tempo information from the reference audio:  
\begin{equation}
    \boldsymbol{F}_{\text{timbre}} = \text{Enc}_{\text{timbre}}(x_{\text{ref}}), \\
    \boldsymbol{F}_{\text{tempo}} = \text{Enc}_{\text{tempo}}(x_{\text{ref}}).
\end{equation}

where $F_{timbre}$ captures the voiceprint-like characteristics of the prompt audio and  
$F_{tempo}$ encodes the overall rhythm information of the prompt audio. These two representations, as depicted in the top-left of Fig.~\ref{flowchart}, are subsequently fused and injected into the model through global conditioning and then frame-level conditioning pathways, thereby enabling control over the style of the generated audio.
\begin{equation}
    \boldsymbol{F}_{style} = \text{MLP}(\boldsymbol{F}_{timbre} + \boldsymbol{F}_{tempo}).
\end{equation}

The fused style representation is injected into the model through global and then frame-level conditioning:  
\begin{equation}
    \begin{split}
    &\boldsymbol{C}_{g} = \boldsymbol{F}_{style} + \text{MLP}(\boldsymbol{F}_{avg\_v} + \boldsymbol{F}_{avg\_t}) + \boldsymbol{t}, \\
    &\boldsymbol{C}_{f} = \text{Upsample}(\text{ConvMLP}(F_{syn})) + \boldsymbol{C}_{g}.
    \end{split}
\end{equation}

\begin{table*}[htbp]
\small
\renewcommand{\arraystretch}{1.2}
\caption{Video-to-audio performance on the VGGSound test set. Metrics include distribution matching (KL\textsubscript{PANNs}, KL\textsubscript{PaSST}), semantic alignment (IB-score), and temporal synchronization (DeSync).}
\vspace{-5pt}
\centering
\begin{tabular}{lcccc}
\toprule[1pt]
\textbf{Method} & \multicolumn{2}{c}{\textbf{Distribution matching}} & \textbf{Semantic align} & \multicolumn{1}{c}{\textbf{Temporal align}} \\ 
\cmidrule(lr{\dimexpr 4\tabcolsep-16pt}){2-3}
\cmidrule(lr{\dimexpr 4\tabcolsep-16pt}){4-4}
\cmidrule(lr{\dimexpr 4\tabcolsep-16pt}){5-5}
& \textbf{KL\textsubscript{PANNs}} $\downarrow$ & \textbf{KL\textsubscript{PaSST}} $\downarrow$ & \textbf{IB-score} $\uparrow$ & \textbf{DeSync} $\downarrow$ \\ \midrule
MMAudio(Vanilla)            & 1.72 & 1.50 & 31.75 & 0.53 
\\
MMAudio w/ Prompt Masking  & 1.71 & 1.54 & 30.08 & 0.50 
\\
AudioIM w/o Style Enc     & 1.70 & 1.49 & 31.21 & 0.51 
\\
AudioIM(ours)              & \textbf{1.65} & \textbf{1.44} & \textbf{31.98} & \textbf{0.49} \\
\bottomrule[1pt]
\end{tabular}
\label{tab:table1}
\end{table*}

\section{Experiment}

\subsection{Dataset}

We fine-tune the ‘L-44.1kHz’ version of MMAudio on the VGGSound \cite{chen2020vggsound} dataset. 
VGGSound is a large-scale audio-visual dataset consisting of approximately 200K 10-second video clips spanning 309 sound classes. The total duration of the dataset is around 550 hours. Following prior work such as ReWaS \cite{jeong2025read} and VATT \cite{liu2024tell}, we utilize the class labels (and their natural language names) as textual inputs. We sample the first 8 seconds of each video-audio pair. The first 3 seconds of audio are designated as the prompt audio, while the subsequent 5 seconds are used as the target segment for generation. For training, we use the official training split approximately 180K training clips. All models are fine-tuned on this training set and evaluated on the official VGG-Sound test set, which contains 50 clips per class for reliable performance benchmarking.

\subsection{Metrics}

For the video-to-audio generation, we evaluate the similarity between the feature distributions of real and generated audio under specific embedding models using distribution matching. In line with the methodology of \cite{liu2023audioldm}, we employ PANNs ($\text{KL}_{\text{PANNs}}$) and PaSST ($\text{KL}_{\text{PaSST}}$) as classifiers to calculate the Kullback–Leibler Divergence (KL). 
In addition, we employ the ImageBind method \cite{girdhar2023imagebind} to compute the cosine similarity between video and audio features, yielding the “IB score”, which reflects the semantic alignment between the input video and the generated audio. To assess audio–video synchronization, we utilize the Synchformer \cite{iashin2024synchformer} to calculate the DeSync score. Specifically, we evaluate synchronization by averaging the results obtained from the first 4.8-second and the last 4.8-second segments of the video–audio pair, thereby measuring the temporal alignment between the generated 5-second audio and the corresponding video.

To evaluate the stylistic similarity between the generated audio and the prompt audio, we perform distribution matching between them using ($\text{KL}_{\text{PANNs}}$) and ($\text{KL}_{\text{PaSST}}$). In addition, human evaluation is conducted to further validate the results(the evaluators received professional training).

\subsection{Experimental Settings}

We evaluated the performance of our proposed method on MMAudio ‘L-44.1kHz’. 
The ‘L-44.1kHz’ version of MMAudio has 7 multi-modal transformer blocks and 14 single-modal transformer blocks, hidden dimension of model is 896. 
We used Adam optimizer with learning rate $1\times 10^{-5}$. 
In the training stage, batch size is 512, the model was trained for 800 steps, about 2 epochs.
To enable classifier-free guidance during inference, we randomly mask away the visual tokens (${F}_{v}$ and ${F}_{syn}$), the text or the prompt audio (${x}_{p}$) with a 10\% probability during training. The masked visual tokens are replaced with learnable tokens ($\varnothing_{v}$ and $\varnothing_{syn}$), while any masked text is replaced with the empty string $\varnothing_{t}$, and any masked prompt audio is replaced with the empty string $\varnothing_{p}$.

\subsection{Experimental Results and Analysis}

\subsubsection{Overall Comparison}

Table~\ref{tab:table1} reports the performance of different models in terms of distribution matching between generated and ground-truth audio, audio–video semantic and temporal alignment. 
Specifically, MMAudio (Vanilla) refers to the original MMAudio ‘L-44.1kHz’ version evaluated on video-to-audio generation. 
MMAudio w/ Prompt Masking denotes the same MMAudio version guided by a 3-second prompt only during inference. 
AudioIM w/o Style Enc represents MMAudio fine-tuned with masked prompts and video–audio conditions, but without the style features. 
AudioIM (ours) further incorporates the style encoders to extract style representations and introduces global and then frame-level conditioning. 
Compared with the MMAudio w/ Prompt Masking method, our approach reduces $\text{KL}_{\text{PANNs}}$, and $\text{KL}_{\text{PaSST}}$ by 0.06, and 0.1, respectively, indicating a closer match between generated and ground-truth audio under prompt audio conditions. 
It also improves the IB-score by 1.9, while reducing DeSync by 0.01, demonstrating enhanced audio quality without compromising content consistency or temporal alignment with the video.
Compared with the MMAudio (Vanilla), our method reduces $\text{KL}_{\text{PANNs}}$, and $\text{KL}_{\text{PaSST}}$ by 0.07 and 0.06, respectively, demonstrating that it does not compromise the authenticity of the generated audio. It also improves the IB-score by 0.23, while reducing DeSync by 0.04, demonstrating that the content consistency and temporal alignment with the video are not compromised.
Experimental results show that incorporating prompt guidance improves the model performance, while successive masked training and the introduction of the style encoders further boost overall performance.

\begin{table}[htbp]
\renewcommand{\arraystretch}{1.2}
\caption{Style similarity evaluation between generated and reference audio on the VGGSound test set. Objective distribution matching metrics and subjective SS-MOS are reported. Subjective scores are averaged over 20 listeners with 95\% confidence intervals.}
\vspace{-5pt}
\centering
\resizebox{\linewidth}{!}{%
\begin{tabular}{lccc}
\toprule[1pt]
\textbf{Method} & \multicolumn{2}{c}{\textbf{Distribution matching}} & \textbf{SS-MOS}\\ 
\cmidrule(lr{\dimexpr 4\tabcolsep-16pt}){2-3}
& \textbf{KL\textsubscript{PANNs}} $\downarrow$ & \textbf{KL\textsubscript{PaSST}} $\downarrow$ & \\ \midrule
MMAudio(Vanilla)            & 1.95 & 1.72 & $3.22\pm0.24$\\
MMAudio w/ Prompt Masking  & 1.89 & 1.71 & $3.59\pm0.25$\\
AudioIM w/o Style Enc     & 1.90 & 1.68 & $3.63\pm0.28$\\
AudioIM(ours)              & \textbf{1.85} & \textbf{1.63} & $\textbf{4.06}\pm0.21$\\
\bottomrule[1pt]
\end{tabular}}
\label{tab:table2}
\end{table}

\subsubsection{Style Similarity}\label{sec:figures}

Table~\ref{tab:table2} reports style similarity between generated and reference audio, including objective distribution matching metrics ($\text{KL}_{\text{PANNs}}$, $\text{KL}_{\text{PaSST}}$) and subjective style similarity mean opinion score (SS-MOS). Lower KL scores indicate closer alignment in timbre and tempo characteristics. SS-MOS is the human evaluation metric that assesses the perceptual similarity of timbre and tempo between generated and reference audio, where higher scores indicate greater perceived style similarity.

Compared with MMAudio (Vanilla), MMAudio w/ Prompt Masking achieves better distribution alignment with the reference audio. AudioIM w/o Style Enc further enhances the distribution matching between the generated and reference audio. Building upon AudioIM w/o Style Enc, AudioIM introduces style feature through global-level conditional injection. With this design, AudioIM achieves state-of-the-art distribution matching scores.
We conduct subjective evaluation using the style similarity (SS-MOS). As additional conditioning components are introduced, style similarity consistently improves. Since MMAudio (Vanilla) does not utilize prompt audio during inference, its SS-MOS serves as a lower-bound baseline. Latent prompt conditioning substantially enhances similarity, while structured style encoders provide further refinement. Importantly, these gains are achieved without degrading semantic alignment or synchronization (Table~\ref{tab:table1}).
These results indicate that our method enables controllable timbre and tempo modeling without significantly degrading the semantic consistency and temporal alignment of the video-to-audio system.

Additionally, during subjective evaluation, we observe that the proposed framework demonstrates stronger style control on non-vocal sound effects, particularly instrumental sounds. In contrast, style modulation for vocal sound effects appears less pronounced. A possible explanation is that the tempo encoder is pretrained primarily on large-scale music data, which may introduce a distributional bias toward rhythmic instrumental patterns.

We further analyze the roles of different conditioning components. When style representations are introduced without latent-level prompt conditioning, the generated audio becomes unstable, suggesting that global style features alone may be insufficient for fine-grained control. This indicates that latent prompt conditioning and global style representations play complementary roles. Empirically, removing either the timbre encoder or the tempo encoder reduces style similarity, suggesting that both components contribute to overall performance.

\subsubsection{Qualitative Case Study}
\begin{figure}[htbp]
  \centering
  \includegraphics[width=\linewidth]{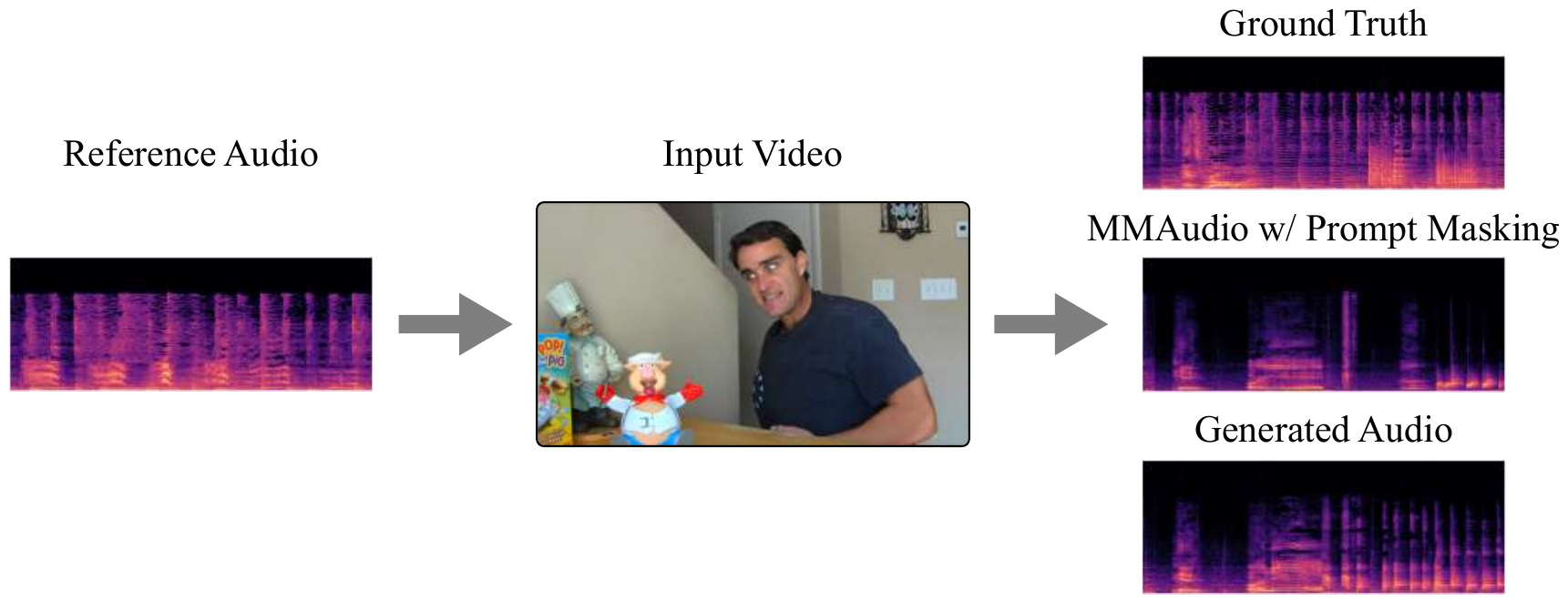}
  \vspace{-5pt}
  \caption{Qualitative comparison of style control. The generated audio reflects the reference tempo and exhibits similar timbre, while remaining consistent with the input video.}
  \label{fig:case_study}
\end{figure}

Figure~\ref{fig:case_study} provides a qualitative comparison. 
The reference spectrogram exhibits periodic vertical structures corresponding to rhythmic patterns. 
Our generated audio shows similar temporal spacing of these structures, whereas prompt masking produces less regular rhythm. 
In terms of timbre, the spectral energy distribution of our result is closer to the reference, which can be further confirmed by listening examples on the demo page. 
These observations are consistent with the style similarity improvements reported in Table~\ref{tab:table2}.
More samples are available online.

\section{CONCLUSIONS}

We presented AudioIM, an attribute-aware video-to-audio generation framework that introduces reference conditioning for style modulation. By combining latent prompt conditioning with dual style encoders for timbre and tempo modeling, the proposed method improves style similarity while preserving semantic alignment and synchronization. Experiments on VGGSound demonstrate that attribute-aware style conditioning can enhance controllability without degrading core video-to-audio performance.

\section{Generative AI Use Disclosure}
During the preparation of this manuscript, the authors used generative AI tools to polish the English language, improve readability, and assist with \LaTeX{} formatting. These tools were not used to generate any scientific claims, experimental results, or significant parts of the manuscript.

\vfill\pagebreak




\bibliographystyle{IEEEtran}
\bibliography{mybib}

\end{document}